\documentstyle[12pt]{article}

\begin{document}

\title{\huge \bf Gauge dependence in the theory of non-linear spacetime 
perturbations}

\author{Sebastiano Sonego\thanks{E-mail: sonego@galileo.sissa.it} {}
and Marco Bruni\thanks{E-mail: bruni@chandra.ap.sissa.it}\\  {\small \em
International School for Advanced Studies, Via Beirut 2-4,
34014 Trieste, Italy}}

\date{August 1997}

\def\theequation{\arabic{section}.\arabic{equation}}
\newtheorem{proposition}{Proposition}
\newtheorem{theorem}[proposition]{Theorem}

\maketitle

\thispagestyle{empty}

\newcommand{\beq}{\begin{equation}}
\newcommand{\eeq}{\end{equation}}
\newcommand{\lab}{\label}
\newcommand{\dd}{{\rm d}}
\newcommand{\ee}{{\rm e}}
\newcommand{\IR}{{\rm I\!R}}
\newcommand{\blambda}{{\mbox{\boldmath$\lambda$}}}

\vspace{2cm}

\begin{abstract}

Diffeomorphism freedom induces a gauge dependence in the theory of 
spacetime perturbations.  We derive a compact formula for gauge 
transformations of perturbations of arbitrary order.  To this end, 
we develop the theory of Taylor expansions for one-parameter 
families (not necessarily groups) of 
diffeomorphisms.  First, we introduce the notion of knight 
diffeomorphism, that generalises the usual concept of flow, and 
prove a Taylor's formula for the action of a knight on a general 
tensor field.  Then, we show that any one-parameter family of 
diffeomorphisms can be approximated by a family of suitable 
knights.  Since in perturbation 
theory the gauge freedom is given by a one-parameter family of 
diffeomorphisms, the expansion of knights is used to derive our 
transformation formula.  The problem of gauge dependence is a 
purely kinematical one, therefore our treatment is valid not only in general 
relativity, but in any spacetime theory.

\end{abstract}

\bigskip\bigskip\bigskip
\bigskip\bigskip\bigskip
\centerline{SISSA--105/97/A}
\bigskip\bigskip\bigskip
\centerline{{\it to appear in Communications in Mathematical Physics}}  
\newpage

\baselineskip=18pt

\setcounter{page}{1}

\section{Introduction}

\setcounter{equation}{0}

In the theory of spacetime perturbations \cite{bi:SW,bi:waldbook,bmms}, 
one usually deals with a family of spacetime models $M_\lambda:=({\cal 
M},\{T_\lambda\})$, where $\cal M$ is a manifold that accounts for the 
topological and differential properties of spacetime, and $\{T_\lambda\}$ 
is a set of fields on $\cal M$, representing its geometrical and physical 
content.  The numerical parameter $\lambda$ that labels the various members 
of the family gives an indication of the `size' of the perturbations,
regarded as deviations of $M_\lambda$ from a background model $M_0$. 
Perturbations are described as additional fields in the background, defined
as $\Delta T^\varphi_\lambda :=\varphi^*_\lambda T_\lambda -T_0$, where 
$\varphi_\lambda : {\cal M}\to {\cal M}$ is a diffeomorphism that provides 
a pairwise identification between points of the perturbed spacetime and of 
the background, and $\varphi_\lambda^*$ denotes the pull-back.  Of course,
such an identification is arbitrary, and this leads to a gauge freedom in the 
definition of perturbations.  Under a  change $\varphi_\lambda\to 
\psi_\lambda$ of the point identification mapping, a perturbation 
transforms as $\Delta T^\varphi_\lambda \to \Delta T^\psi_\lambda$, with 
\beq \Delta T^\psi_\lambda=\Phi_\lambda^* \Delta T^\varphi_\lambda +
\left(\Phi^*_\lambda T_0-T_0\right)\;,
\lab{gauge}\eeq
where $\Phi_\lambda:=\varphi_\lambda^{-1}\circ\psi_\lambda$ is a 
diffeomorphism on $\cal M$.

In the perturbative approach, one tries to approximate $T_\lambda$ expressing 
$\Delta T^\varphi_\lambda$ as a series,
\beq \Delta T^\varphi_\lambda=\sum_{k=1}^{n-1}{\lambda^k\over
k!}\,\delta^k T^\varphi+O(\lambda^n)\;,
\lab{exp}\eeq
where $n$ is the order of differentiability with respect to $\lambda$ of
$\Delta T^\varphi_\lambda$, and then solving iteratively the field equations
for the various terms $\delta^k T^\varphi$.  It is then important to know 
how the latter transform under a change of gauge.  Until very recently, 
only the first order terms, $\delta^1 T^\varphi$, have been considered; 
in this case, it is well-known  that the representations of a 
perturbation in two different gauges
differ  just by a Lie derivative of the background quantity $T_0$ 
\cite{bi:SW}.  However, non-linear perturbations are now becoming a 
valuable tool of investigation in  black hole and  gravitational wave 
physics \cite{bi:gleiseretal}, as well as in cosmology \cite{bi:cosmo}.  
Their behaviour under gauge transformations  can be derived by 
Taylor-expanding (\ref{gauge}) with respect to $\lambda$.  

This apparently straightforward procedure presents a difficulty, though.  
Even if one chooses, as usual, point identification maps that are one-parameter
groups with respect to $\lambda$, the family of diffeomorphisms 
$\Phi_\lambda$ is {\em not\/} a one-parameter group \cite{bmms}, i.e., it
does not correspond to a flow on $\cal M$.  While flows on manifolds
are well understood and widely discussed in the literature, more general
one-parameter families of diffeomorphisms are not.  Only some fragmentary
statements about them can be found in a few papers 
\cite{taub,schutz1,gl,fw}.  Therefore, in order to extract from 
(\ref{gauge}) the relationship between
$\delta^k T^\varphi$ and $\delta^k T^\psi$, one must first develop the
theory of Taylor expansions for general one-parameter {\em families\/} of
diffeomorphisms, not necessarily forming a local group.

The purpose of the present article is to provide the mathematical framework 
needed for this purpose.  Roughly, the discussion generalises section 2 of
reference \cite{bmms} from  the analytic to the $C^n$ case, but we also 
derive here a compact formula that gives directly the gauge 
transformation to an arbitrary order $k$.  The paper is organised 
as follows.  In the next section we define particular combinations of flows
that we dub knight diffeomorphisms, and present our main result (Theorem
\ref{theorem1}).  This establishes that arbitrary one-parameter families of
diffeomorphisms can be approximated by families of knights, so that all one needs is a
suitable expression for the Taylor expansion of knights, which is derived in
section 3.  Then, Theorem \ref{theorem1} is proved in section 4.  Section 5
contains the application to (\ref{gauge}), i.e., our formula 
(\ref{gtransf}) and some concluding remarks.

In the following, we shall work on a finite-dimensional 
manifold $\cal M$, smooth enough for all the statements below to 
make sense.  In order to avoid cumbersome talking about 
neighbourhoods, we shall often suppose that maps are globally
defined.  This assumption simplifies the discussion, without 
altering the results significantly.  Also, we specify the class 
of differentiability of an object only when it is really needed.  
Finally, let us recall that a one-parameter family of diffeomorphisms 
of $\cal M$ is a differentiable mapping $\Phi:{\cal D}\to{\cal 
M}$, with $\cal D$  an open subset of ${\rm I\!R}\times{\cal M}$ containing
$\{0\}\times{\cal M}$, and $\Phi(0,p)=p$, $\forall p\in{\cal M}$.  As we
have already been doing, we shall write, following the common usage,
$\Phi_\lambda(p):= \Phi(\lambda,p)$, for any $(\lambda,p)\in{\cal D}$.


\section{Knight diffeomorphisms}

\setcounter{equation}{0}

Let $\phi^{(1)}:{\cal D}_1\to{\cal 
M}, \ldots, \phi^{(k)}:{\cal D}_k\to{\cal M}$ be flows on 
${\cal M}$, generated by the vector fields $\xi_1,\ldots,\xi_k$, 
respectively.  We can combine $\phi^{(1)},\ldots,\phi^{(k)}$ to 
define a new one-parameter family of diffeomorphisms $\Psi:{\cal 
D}\to{\cal M}$, with $\cal D$ a suitable open subset of $\IR\times 
{\cal M}$ containing $\{0\}\times {\cal M}$, whose action is given 
by
\beq \Psi_\lambda:=\phi^{(k)}_{\lambda^k/k!}
\circ\cdots\circ\phi^{(2)}_{\lambda^2/2}
\circ\phi^{(1)}_\lambda\;.
\lab{eq:knightn}\eeq
Thus, $\Psi_\lambda$ displaces a point of ${\cal M}$ a parameter 
interval $\lambda$ along the integral curve of $\xi_1$, then an 
interval $\lambda^2/2$ along the integral curve of $\xi_2$, and 
so on (see Fig.\ 1 for the case $k=2$). For this reason, we shall 
call $\Psi_{\lambda}$, with a chess-inspired terminology, a {\em knight 
diffeomorphism of rank $k$\/} or, more shortly, a {\em knight\/}.  
The vector fields $\xi_1,\ldots,\xi_k$ will be called the {\em 
generators\/} of $\Psi$.

The utility of knights stems from the fact that any $C^n$ one-parameter 
family $\Phi$ of diffeomorphisms can always be approximated by a 
family $\Psi$ of knights of rank $n-1$, as shown by the following

\begin{theorem}  Let $\Phi:{\cal D}\to {\cal M}$ be a $C^n$
one-parameter family of diffeomorphisms.  Then $\exists$
$\phi^{(1)},\ldots,\phi^{(n-1)}$, flows on $\cal M$ such
that, up to the order $\lambda^n$, the action of $\Phi_\lambda$ is 
equivalent to the one of the $C^n$ knight
\beq \Psi_\lambda=\phi^{(n-1)}_{\lambda^{n-1}/(n-1)!}\circ\cdots
\circ\phi^{(2)}_{\lambda^2/2}\circ\phi^{(1)}_\lambda\;.
\lab{theorem}\eeq
\lab{theorem1}\end{theorem}

This result allows one to use knights in order to investigate 
many properties of arbitrary diffeomorphisms.  In a sense, knights play 
among the one-parameter families of diffeomorphisms of $\cal M$ the same 
crucial role that polynomials play for functions of a real variable.  We 
postpone the proof of Theorem \ref{theorem1} to section 4, after we have 
established some preliminary results.


\section{Taylor expansion of flows and knights}

\setcounter{equation}{0}

It is easy to generalise the usual Taylor's expansions on $\IR^m$ 
\cite{bi:ycm} to the case of a flow acting on a manifold:

\begin{proposition}  Let $\phi:{\cal D}\to{\cal M}$ be a flow  generated 
by the vector field $\xi$, and $T$ a tensor  field such that 
$\phi^*_{\lambda} T$ is
a (tensor-valued) function of $\lambda$ of class $C^n$.  Then, 
$\phi^*_{\lambda} T$ can be expanded around $\lambda=0$ as 
\beq \phi^*_{\lambda} T=\sum^{n-1}_{l=0}
\,\frac{\lambda^l}{l!}\,\pounds^l_\xi T+\lambda^n R_{\lambda}^{(n)}T\;,
\lab{lemma1}\eeq where $\pounds_\xi$ is the Lie derivative along the flow
$\phi$, and
$R_{\lambda}^{(n)}$ is a linear map whose action on $T$ is given by
\beq R_{\lambda}^{(n)}T={1\over (n-1)!}\int_0^1\dd t\,\left(1-t\right)^{n-1}
\pounds_\xi^n\phi^\ast_{t\lambda}T\;.
\lab{Rn}\eeq
\lab{Lemma2}\end{proposition}

This proposition has the important consequence that, for a tensor 
field $T$ and a flow $\phi$ such that $\phi^*_\lambda T$ is $C^n$, 
one can approximate $\phi^*_\lambda T$, to order $n-1$, by the polinomial
\[ \sum_{l=0}^{n-1}{\lambda^l\over l!}\,\pounds_\xi^l T\;. \]   
This follows from the property 
\beq \lim_{\lambda\rightarrow 0}R_\lambda^{(n)} T={1\over n!}\,
\pounds_\xi^n T\;,
\lab{limRn}\eeq 
which implies that, for $\lambda\rightarrow 0$, the remainder 
$\lambda^nR_\lambda^{(n)}T$ is $O(\lambda^n)$.\footnote{Actually,
this result holds also for the weaker case in which
$\phi^*_\lambda T$ is $C^{n-}$ (i.e., it is of class $C^{n-1}$ with a locally
Lipschitzian $(n-1)$th derivative).  However, under these conditions one
does not have an explicit expression, like (\ref{Rn}), for the 
remainder.}

The proof of Proposition \ref{Lemma2} is rather straightforward 
and can be omitted.  We only wish to point out that it relies heavily on the 
property that $\phi_\lambda$ forms a one-parameter group:
$\phi_{\sigma+\lambda}=\phi_\sigma \circ \phi_\lambda$.  It is evident 
from (\ref{eq:knightn}) that for knights one has, in general, 
$\Psi_\sigma\circ\Psi_\lambda\neq\Psi_{\sigma+\lambda}$, and
$\Psi_\lambda^{-1}\neq\Psi_{-\lambda}$.  Thus, equation 
(\ref{lemma1}) cannot be applied if we want to expand in
$\lambda$ the pull-back $\Psi_{\lambda}^* T$ of a tensor field $T$
defined on ${\cal M}$.  The ultimate reason for this, is that 
a family of knights does not form a group, except under very special 
conditions, as shown by the following

\begin{theorem}  Let $\Psi:{\cal D}\to{\cal M}$ be a family of knight
diffeomorphisms of rank $k$, with generators $\xi_1,\ldots,\xi_k$. 
$\Psi$ forms a group iff there exists a vector field $\xi$, and
numerical coefficients $\alpha_l$, with $1\leq l\leq k$, such that
$\xi_l=\alpha_l\xi$, $\forall\,l$.  In this case, under the
reparametrisation $\lambda\to\bar{\lambda}:= f(\lambda)$, with 
\beq f(\lambda):= \sum_{l=1}^k\alpha_l\lambda^l/l! \;,
\eeq
$\Psi$ reduces to a flow in the canonical form.
\end{theorem}

\noindent {\em Proof.} Let us first show that $\xi_l=\alpha_l\xi$ is a
sufficient condition for $\Psi$ to form a group.  Let $\phi$ be the
flow generated by $\xi$.  Then $\phi_\sigma^{(l)}=
\phi_{\alpha_l\sigma}$, and we have $\Psi_\lambda=
\phi_{\alpha_k\lambda^k/k!}\circ\cdots
\circ\phi_{\alpha_1\lambda}=\phi_{\bar{\lambda}}$.  Thus, (i)
$\Psi_\sigma\circ\Psi_\lambda=
\phi_{\bar{\sigma}}\circ\phi_{\bar{\lambda}}=
\phi_{\bar{\sigma}+\bar{\lambda}}=
\phi_{\bar{\tau}}=\Psi_\tau$, with
$\tau=f^{-1}(\bar{\sigma}+\bar{\lambda})$, and (ii)
$\Psi_\lambda^{-1}=\phi^{-1}_{\bar{\lambda}}=
\phi_{-\bar{\lambda}}=\Psi_\rho$, with
$\rho=f^{-1}(-\bar{\lambda})$.

To prove the reverse implication, let us suppose that $\Psi$ form a
group.  Let $p$ be an arbitrary point of $\cal M$, and define the set
${\cal C}_p:=\{\Psi_\lambda(p)|\lambda\in I_p\}\subset {\cal M}$, where
$I_p\ni 0$ is an open interval of $\IR$ such that $I_p\times\{p\}
\subset{\cal D}$.  Obviously, ${\cal C}_p$ is a one-dimensional 
submanifold of $\cal M$ (to see this, it is sufficient to consider a
chart on ${\cal C}_p$ where $\lambda$ itself is the coordinate).  Let us
now consider another arbitrary point $q\in{\cal C}_p$, and ask whether
it is possible that ${\cal C}_q\neq{\cal C}_p$.  If it were so, there
would be some $\sigma\in I_q$ such that $\Psi_\sigma(q)\neq\Psi_\tau(p)$,
$\forall\tau\in I_p$.  But since $q=\Psi_\lambda(p)$, for some $\lambda\in
I_p$, and $p$ is arbitrary, this would mean that, for some $\lambda$ and
$\sigma$, one cannot find a $\tau$ such that $\Psi_\sigma\circ\Psi_\lambda
=\Psi_\tau$, which would contradict the hypothesis that $\Psi$ forms a
group.  Thus, each point of $\cal M$ belongs to one, and only one,
one-dimensional submanifold constructed using $\Psi$ as above.  The set of
these submanifolds becomes a congruence of curves simply by suitably
parametrising them; this, in turn, defines a flow $\phi$ and a vector field
$\xi$.  Thus, if $\Psi$ forms a group, it can be written as
$\Psi_\lambda=\phi_{\bar{\lambda}}$, for some suitable parameter
$\bar{\lambda}$.

In the particular case of a knight, this condition can be rewritten, using
(\ref{eq:knightn}), as
\beq \phi^{(1)}_\lambda\circ\phi_{-\bar{\lambda}}=
\phi^{(2)}_{-\lambda^2/2}\circ\cdots
\circ\phi^{(k)}_{-\lambda^k/k!}\;.
\lab{porcoqua}\eeq
Assuming $\phi$ and the various $\phi^{(l)}$ to be at least of class 
$C^{2-}$ (which is a natural requirement, if one wants them to be uniquely
determined by the respective vector fields), we can apply (\ref{lemma1})
to (\ref{porcoqua}) and get, for an arbitrary tensor field $T$,
\beq \left(\pounds_{\xi_1}-
{\bar{\lambda}\over\lambda}\,\pounds_\xi\right)T=O(\lambda)\;.
\eeq
This implies that $\exists\alpha_1\in\IR$ such that 
$\bar{\lambda}=\alpha_1\lambda+f_2(\lambda)$, with
$f_2(\lambda)=O(\lambda^2)$, together with
$\xi_1=\alpha_1\xi$.  Substituting into (\ref{porcoqua}) and applying
again (\ref{lemma1}), we find
\beq \left(\pounds_{\xi_2}-
{f_2({\lambda})\over\lambda^2/2}\,\pounds_\xi\right)T=O(\lambda)\;.
\eeq
Thus, we have also that $\exists\alpha_2\in\IR$ such that 
$f_2(\lambda)=\alpha_2\lambda^2/2+f_3(\lambda)$, with
$f_3(\lambda)=O(\lambda^3)$, and $\xi_2=\alpha_2\xi$.  Iterating this
procedure,  one shows that $\xi_l=\alpha_l\xi$, $\forall l\leq
k$.\hfill$\Box$\\

It is clear from the proof given above that the failure of $\Psi$ to
form a group is also related to the following circumstance.  For any
$p\in{\cal M}$, one can define a curve $u_p:I_p\to{\cal M}$ by
$u_p(\lambda):=\Psi_\lambda(p)$.  However, these curves do not form a
congruence on $\cal M$.  For the point $u_p(\lambda)$, say, belongs 
not only to the image of the curve $u_p$, but also to the one of 
$u_{u_p(\lambda)}$, which differs from $u_p$ when at least
one of the $\xi_l$ is not collinear with $\xi_1$, since
$u_{u_p(\lambda)}(\sigma) =\Psi_\sigma\circ\Psi_\lambda(p)\neq
\Psi_{\lambda+\sigma}(p)= u_p(\lambda+\sigma)$.  Thus, the fundamental
property of a congruence, that each point of $\cal M$ lies on the image
of one, and only one, curve, is violated.

Let us now turn to the problem of Taylor-expanding $\Psi_{\lambda}^* 
T$.  Although (\ref{lemma1}) cannot be used straightforwardly for 
this purpose, one can apply it repeatedly to
$\Psi_{\lambda}^* T=\phi^{(1)*}_{\lambda}\phi^{(2)*}_{\lambda^2/2}
\cdots \phi^{(k)*}_{\lambda^k/k!} T$, and get the following

\begin{proposition} Let $\Psi$ be  a one-parameter family of knight
diffeomorphisms of rank $k$, and $T$ a tensor field such that
$\Psi_{\lambda}^* T$ is  of class  $C^n$.
 Then  $\Psi_{\lambda}^* T$ can be expanded around $\lambda=0$ as 
\beq \Psi_{\lambda}^* T=\sum_{l=0}^{n-1}{\lambda^l\over l!}\sum_{J_l}
{l!\over 2^{j_2}\cdots n!^{j_n} j_1!j_2!\cdots
j_n!}\,\pounds^{j_1}_{\xi_1}
\cdots\pounds^{j_n}_{\xi_n} T+\lambda^nR^{(n)}_\lambda T\;,
\lab{lemma2}\eeq 
where $J_l:=\{(j_1,\ldots,j_n)\in {\rm I\!N}^n|\,
\sum_{i=1}^n i\,j_i=l\}$ defines the set of indices over which one has to sum
in order to obtain the $l$-th order term, and
$R^{(n)}_\lambda T$ is a remainder with a finite limit as 
$\lambda\rightarrow 0$.
\end{proposition}

The geometrical meaning of (\ref{lemma2}) is particularly 
clear in a chart.  Let us consider the special case in
which the tensor $T$ is just one of the coordinate functions on $\cal M$,
$x^\mu$.  We have then, since $\Psi^*_{\lambda}x^\mu(p)=
x^\mu(\Psi_\lambda(p))$, the action of an `infinitesimal point
transformation,' that reads, to second order in $\lambda$,
\beq \tilde{x}^\mu=x^\mu +\lambda\,\xi_1^\mu+{\lambda^2\over
2}\,\left({\xi_1^\mu}_{,\nu}\xi_1^\nu +\xi_2^\mu\right)+O(\lambda^3)\;,
\lab{lemma2coord}\eeq
where we have denoted $x^\mu(p)$ simply by $x^\mu$, and
$x^\mu(\Psi_\lambda(p))$ by $\tilde{x}^\mu$.  Equation 
(\ref{lemma2coord}) is represented pictorially in Fig.\ 2.  
The effect of $\phi^{(2)}$ (and of higher order $\phi$'s) is 
to correct the action of the simple flow $\phi^{(1)}$.

Finally, let us notice that since each element of $J_l$ has $j_i\equiv 
0$, $\forall i>l$, the sum on the right hand side of (\ref{lemma2})
only involves the Lie derivatives along the vectors $\xi_l$ with $l\leq
n-1$.  Thus, as far as Taylor expansions are concerned, only  knights of rank
lower than their degree of differentiability are really relevant.


\section{Proof of Theorem\ 1}

\setcounter{equation}{0}

If $\varphi$ and $\psi$ are two diffeomorphisms of $\cal M$ such 
that $\varphi^*f=\psi^*f$ for every function $f$, it follows that 
$\varphi\equiv\psi$, as it is easy to see in a chart.  Thus, in 
order to show that a family of knights $\Psi$ approximates
any one-parameter family of diffeomorphisms $\Phi$ up to the $n$-th 
order, it is sufficient to prove 
that $\Psi^*_\lambda f$ and $\Phi^*_\lambda f$ differ by a function 
that is $O(\lambda^n)$, $\forall f$.  Let us therefore consider the 
action of $\Phi_\lambda$ 
on an arbitrary sufficiently smooth function $f:{\cal M}\to{\rm 
I\!R}$.  The Taylor expansion of $\Phi^*_\lambda f$ gives \cite{bi:ycm} 
\beq \Phi_{\lambda}^* f=\sum_{l=0}^{n-1}{\lambda^l\over
l!}\,\left.{{\rm d}^l~\over{\rm d}\lambda^l}\right|_0\Phi^*_{\lambda}
f+\lambda^nR_\lambda^{(n)}f\;,
\lab{ostia}\eeq
with
\beq R_\lambda^{(n)}f={1\over (n-1)!}\int_0^1\dd t\, \left(1-t\right)^{n-1}
\left.{{\rm d}^n~\over{\rm
d}\lambda'^n}\right|_{t\lambda}\Phi^*_{\lambda'} f\;.
\eeq
Let us define $n-1$ linear differential operators ${\cal
L}_1,\ldots,{\cal L}_{n-1}$ through the recursive formula  
\beq {\cal L}_lf:=\left.{{\rm d}^l~\over{\rm d}
\lambda^l}\right|_0\Phi_{\lambda}^* f-\sum_{J'_l} {l!\over 2^{j_2}\cdots
(l-1)!^{j_{l-1}} j_1!j_2!\cdots
j_{l-1}!}\,{\cal L}^{j_1}_1{\cal L}^{j_2}_2
\cdots{\cal L}^{j_{l-1}}_{l-1}f\;,
\lab{kjy}\eeq
where $J_1'\equiv \emptyset$ and, for $l>1$, 
$J'_l:=\{(j_1,\ldots,j_{l-1})\in 
{\rm I\!N}^{l-1}|\, \sum_{i=1}^{l-1}i\,j_i=l\}$.  Since ${\cal
L}_1,\ldots,{\cal L}_{n-1}$ satisfy Leibniz's rule (see Appendix), they
are derivatives, and we can thus define $n-1$ vector fields
$\xi_1,\ldots,\xi_{n-1}$ by requiring that, for any $C^1$ function $f$,   
$\pounds_{\xi_l}f:={\cal L}_l f$.  Now, if
$\Psi_\lambda$ is the knight of rank $n-1$ generated by
$\xi_1,\ldots,\xi_{n-1}$ as in (\ref{theorem}), we can combine (\ref{ostia}),
(\ref{kjy}), and (\ref{lemma2}) to get
\beq \Phi^*_\lambda f=\Psi^*_\lambda f+\lambda^n\Delta^{(n)}_\lambda f\;,
\eeq
where $\Delta^{(n)}_\lambda f$ is $O(\lambda^0)$.  This completes the 
proof.\hfill$\Box$\\


\section{Gauge transformation and conclusions}

\setcounter{equation}{0}

In the previous sections we have presented the theory of Taylor's expansions
for one-parameter families of diffeomorphisms on a manifold $\cal M$.  Taking 
the simple case of a flow as our basic element, we have first defined the 
notion of knights, and then shown that an arbitrary one-parameter family of 
diffeomorphisms can always be approximated by a family of knights of a
suitable rank.  We can now return to the problem stated in the introduction,
of finding the relationship between the $k$th order perturbations of a
tensor $T_\lambda$ in two gauges $\varphi_\lambda$ and $\psi_\lambda$.

Let $n$ be the lowest order of differentiability of the objects contained 
in (\ref{gauge}).  It follows from Theorem \ref{theorem1} that the action of
$\Phi_\lambda$ is equivalent, up to the order $\lambda^n$, with the one of a knight
$\Psi_\lambda$, constructed as in (\ref{theorem}).  Therefore, we can expand
(\ref{gauge}) using (\ref{lemma2}), and find, $\forall k<n$,
\beq \delta^k T^\psi=\sum_{l=0}^k{k!\over 
(k-l)!} \sum_{J_l} {1\over 2^{j_2}\cdots k!^{j_k}j_1!\cdots j_k!}\, 
\pounds_{\xi_1}^{j_1}\cdots 
\pounds_{\xi_k}^{j_k}\delta^{k-l}T^\varphi\;,
\label{gtransf}\eeq
where the various quantities are defined according to (\ref{exp}), 
and $\delta^0 T^\varphi:=T_0$. Equation (\ref{gtransf}) 
gives a complete description of the gauge behaviour of 
perturbations at an arbitrary order.  Among other applications, 
it allows one to obtain easily the conditions for the gauge 
invariance of perturbations to $k$th order; this problem has
been discussed in some detail in reference \cite{bmms}.  Since 
the problem of gauge dependence is purely kinematical, 
(\ref{gtransf}) is valid not only in general relativity, but in any 
geometrical theory of spacetime.

Of course, our treatment can be easily generalised in several ways.  For
instance, it may happen that the perturbations are characterised by several
parameters $\lambda_1,\ldots,\lambda_{\scriptscriptstyle N}$
\cite{bi:waldbook}, so that one is dealing with a $N$-parameter family of
spacetime models $M_{(\lambda_1,\ldots,\lambda_{\scriptscriptstyle N})}$ 
that differ from  the background $M_{(0,\ldots,0)}$.  Correspondingly, gauge
transformations  are associated  to the action of a $N$-parameter family of
diffeomorphisms $\Phi:{\cal D}\to{\cal M}$, where $\cal D$ is an open subset
of $\IR^N\times {\cal M}$ containing $\{(0,\ldots,0)\}\times{\cal M}$, and
$\Phi((0,\ldots,0),p)=p$, $\forall p\in{\cal M}$.  One can then ask several
questions about such an extension of the theory discussed in the present
paper.  However, we leave this topic for future investigations.


\section*{Acknowledgements}

We are grateful to Professor Dennis W.\ Sciama for hospitality at the
Astrophysics Sector of SISSA, and to an anonimous referee for stimulating 
several improvements in the presentation.  MB thanks INFN for financial
support.


\section*{Appendix:  Proof that the operators ${\cal L}_l$ satisfy the
Leibniz rule}
\def\theequation{A.\arabic{equation}}
\setcounter{equation}{0}

Since the operators ${\cal L}_l$ are linear, the Leibniz rule is equivalent
to the condition ${\cal L}_l f^2=2f{\cal L}_l f$, for any $C^1$ function
$f$.  This property can be established for any $l$ by induction.  It
trivially holds for $l=1$, so there exists a vector field $\xi_1$ such
that ${\cal L}_1f=\pounds_{\xi_1}f$, $\forall f$.  Let us suppose that this
is true up to $l-1$, so that there are $l-1$ vector fields
$\xi_1,\ldots,\xi_{l-1}$ such that ${\cal L}_kf=\pounds_{\xi_k}f$, $\forall
k\leq l-1$ and $\forall f$.  Then we must prove that
\beq 2f{\cal L}_l f=\left.{{\rm d}^l~\over{\rm d}
\lambda^l}\right|_0(\Phi_{\lambda}^* f)^2-\sum_{J'_l} {l!\over 2^{j_2}\cdots
(l-1)!^{j_{l-1}} j_1!j_2!\cdots
j_{l-1}!}\,\pounds^{j_1}_{\xi_1}\pounds^{j_2}_{\xi_2}
\cdots\pounds^{j_{l-1}}_{\xi_{l-1}}f^2\;.
\lab{show}\eeq
Recalling (\ref{kjy}), we have 
\begin{eqnarray} \lefteqn{\left.{{\rm d}^l~\over{\rm d}
\lambda^l}\right|_0(\Phi_{\lambda}^* f)^2=\sum_{k=0}^l {l \choose k} 
\left.{{\rm d}^k~\over{\rm d}
\lambda^k}\right|_0(\Phi_{\lambda}^* f)\,\left.{{\rm d}^{l-k}~\over{\rm d}
\lambda^{l-k}}\right|_0(\Phi_{\lambda}^* f)}\nonumber\\
&&=2f{\cal L}_l f+2f\sum_{J'_l} {l!\over 2^{j_2}\cdots
(l-1)!^{j_{l-1}} j_1!j_2!\cdots
j_{l-1}!}\,\pounds^{j_1}_{\xi_1}
\cdots\pounds^{j_{l-1}}_{\xi_{l-1}}f\nonumber\\
&&+\sum_{k=1}^{l-1}{l\choose
k}\left(\sum_{J_k}{k!\over 2^{j_2}\cdots k!^{j_k}
j_1!j_2!\cdots j_k!}\,\pounds^{j_1}_{\xi_1}
\cdots\pounds^{j_k}_{\xi_k}f\right) \nonumber\\
&&\left(\sum_{J_{l-k}}{(l-k)!\over 2^{j_2}\cdots
(l-k)!^{j_{l-k}} j_1!j_2!\cdots
j_{l-k}!}\,\pounds^{j_1}_{\xi_1}
\cdots\pounds^{j_{l-k}}_{\xi_{l-k}}f\right);
\end{eqnarray}
therefore, (\ref{show}) is satisfied iff 
\begin{eqnarray}
\lefteqn{\sum_{J'_l} {l!\over 2^{j_2}\cdots (l-1)!^{j_{l-1}} j_1!j_2!\cdots
j_{l-1}!}\,
\pounds^{j_1}_{\xi_1}\cdots\pounds^{j_{l-1}}_{\xi_{l-1}}f^2} \nonumber\\
&&=2f\sum_{J'_l} {l!\over 2^{j_2}\cdots
(l-1)!^{j_{l-1}} j_1!j_2!\cdots
j_{l-1}!}\,\pounds^{j_1}_{\xi_1}
\cdots\pounds^{j_{l-1}}_{\xi_{l-1}}f\nonumber\\
&&+\sum_{k=1}^{l-1}{l\choose
k}\left(\sum_{J_k}{k!\over 2^{j_2}\cdots k!^{j_k}
j_1!j_2!\cdots j_k!}\,\pounds^{j_1}_{\xi_1}
\cdots\pounds^{j_k}_{\xi_k}f\right) \nonumber\\
&&\left(\sum_{J_{l-k}}{(l-k)!\over 2^{j_2}\cdots
(l-k)!^{j_{l-k}} j_1!j_2!\cdots
j_{l-k}!}\,\pounds^{j_1}_{\xi_1}
\cdots\pounds^{j_{l-k}}_{\xi_{l-k}}f\right),
\lab{gu}\end{eqnarray}
for any $f$ and for any choice of the vector fields
$\xi_1,\ldots,\xi_{l-1}$.  This relationship could be proved by brute
force.  However, it is easier to follow an alternative path.  Let us
consider a knight $\Psi_\lambda$ of rank $l$, generated by the vectors
$\xi_1,\ldots,\xi_{l-1}$, and by a new arbitrary vector $\xi_l$.  Then one 
can compute 
\[ \left.{{\rm d}^l~\over{\rm d} \lambda^l}\right|_0 (\Psi_{\lambda}^*
f)^2 \] 
using (\ref{lemma2}), from which (\ref{gu}) follows straightforwardly.



\newpage

\section*{Figure captions}

\begin{description}

\item[{\rm Fig.\ 1}]
The action of a knight diffeomorphism $\Psi_\lambda$ of rank 2 generated 
by $\xi_1$ and $\xi_2$. Solid lines: integral curves of $\xi_1$. Dashed
lines: integral curves of $\xi_2$.  The parameter lapse between $p$ and 
$\phi^{(1)}_\lambda (p)$ is $\lambda$, and that from $\phi^{(1)}_\lambda 
(p)$ to $\Psi_\lambda(p)$ is $\lambda^2/2$.

\item[{\rm Fig.\ 2}]
The action of a knight diffeomorphism of rank two,
represented in a chart to order $\lambda^2$.

\end{description}

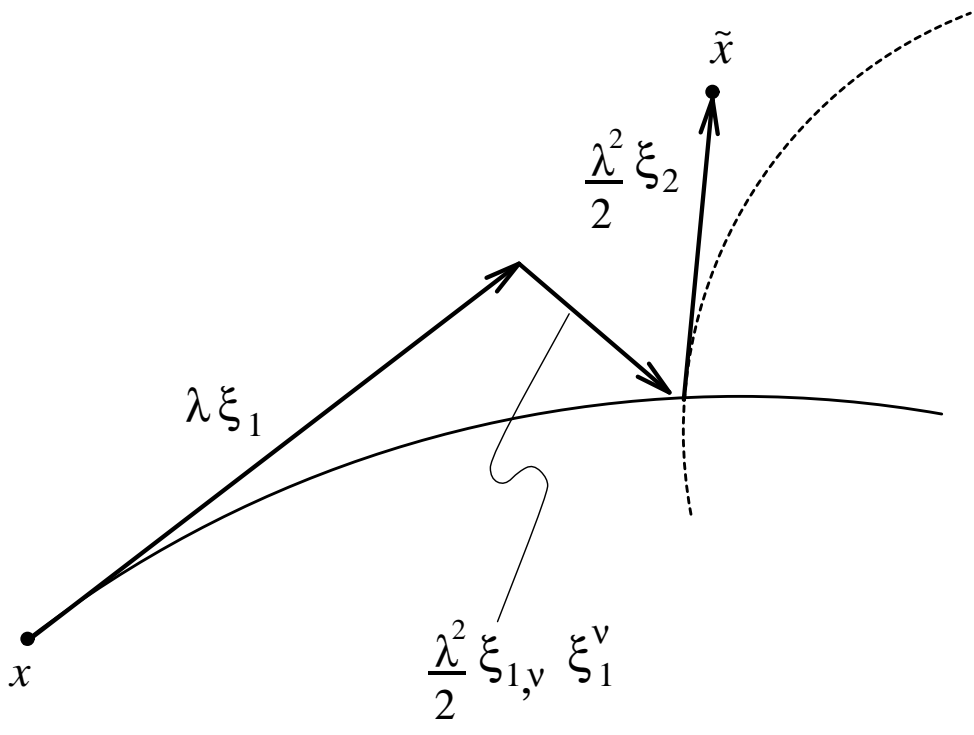
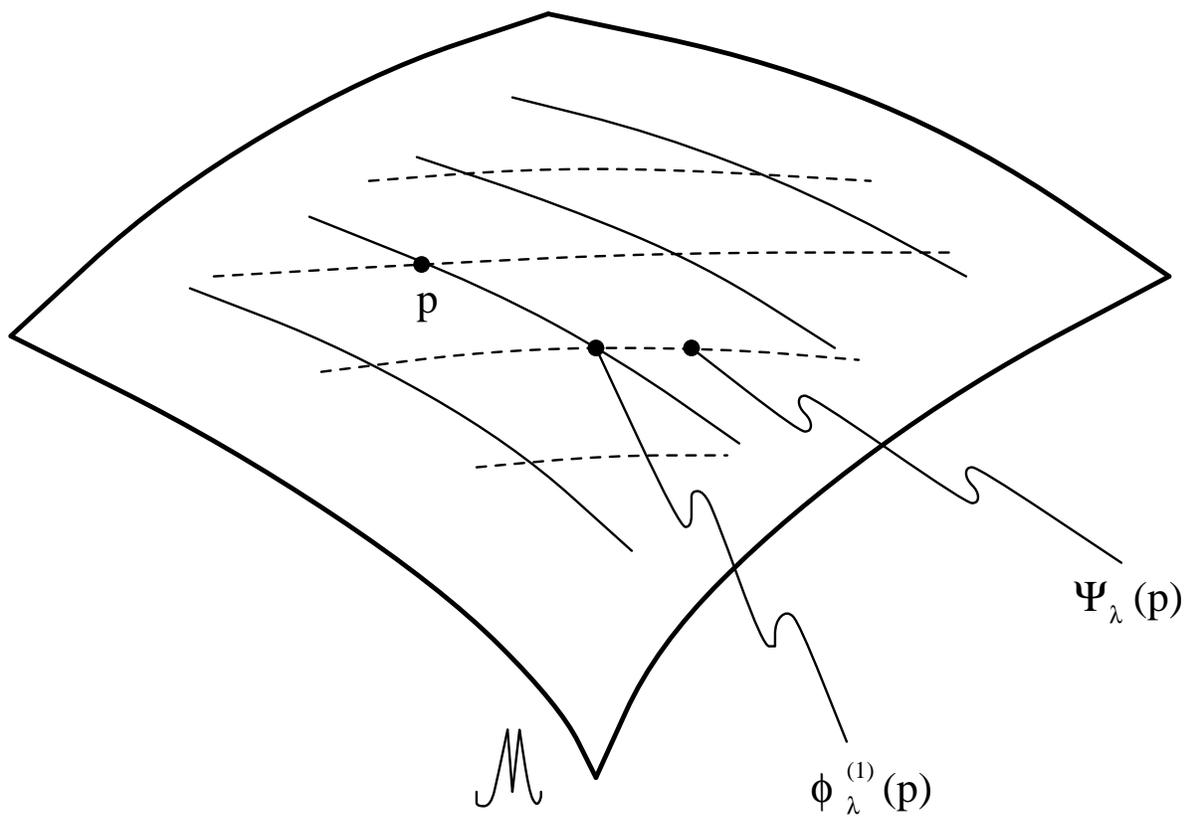
\end{document}